\pdfoutput=1
\documentclass[prologue,table,svgnames,sigconf,nonacm]{acmart}

\usepackage{tugraz_defaults}
\renewcommand{\paragraph}[1]{\noindent\textbf{#1.}\ }

\begin{document}

\newcommand{\DefenseName}{ConTExT\xspace}
\newcommand{\DefenseNameLight}{ConTExT-light\xspace}

\title{\DefenseName: Leakage-Free Transient Execution}

\author{Michael Schwarz}
\affiliation{\institution{Graz University of Technology}}
\email{michael.schwarz@iaik.tugraz.at}

\author{Robert Schilling}
\affiliation{\institution{Graz University of Technology}}
\email{robert.schilling@iaik.tugraz.at}

\author{Florian Kargl}
\affiliation{\institution{Graz University of Technology}}
\email{florian.kargl@student.tugraz.at}

\author{Moritz Lipp}
\affiliation{\institution{Graz University of Technology}}
\email{moritz.lipp@iaik.tugraz.at}

\author{Claudio Canella}
\affiliation{\institution{Graz University of Technology}}
\email{claudio.canella@iaik.tugraz.at}

\author{Daniel Gruss}
\affiliation{\institution{Graz University of Technology}}
\email{daniel.gruss@iaik.tugraz.at}

\renewcommand{\shortauthors}{Schwarz~\etal}

\newcommand{\Spectre}{Spectre\xspace}
\newcommand{\SpectreRetToSpec}{ret2spec\xspace}
\newcommand{\SpectreRSB}{SpectreRSB\xspace}
\newcommand{\NetSpectre}{NetSpectre\space}

\newcommand{\Meltdown}{Meltdown\xspace}
\newcommand{\Foreshadow}{Foreshadow\xspace}
\newcommand{\ForeshadowNG}{Foreshadow-NG\xspace}
\newcommand{\LazyFP}{Lazy FP\xspace}

\newcommand{\safestack}{SafeStack\xspace}

\acmDOI{XX.XXX/XXX_X}

\acmISBN{XXX-XXXX-XX-XXX/XX/XX}

\acmConference[CCS'19]{ACM Conference on Computer and Communications Security}{November 11--15, 2019}{London, UK}
\acmBooktitle{CCS'19: ACM Conference on Computer and Communications Security, November 11--15, 2019, London, UK}
\acmYear{2019}
\copyrightyear{2019}
\acmPrice{15.00}


\begin{abstract}
Out-of-order execution and speculative execution are among the biggest contributors to performance and efficiency of modern processors.
However, they are inconsiderate, leaking secret data during the transient execution of instructions.
Many solutions have been proposed against transient execution attacks.
However, they do not eliminate the leakage entirely or introduce unacceptable performance penalties. 

In this paper, we propose \DefenseName, a \textbf{Con}siderate \textbf{T}ransient \textbf{Ex}ecution \textbf{T}echnique.
The basic idea of \DefenseName is that \textbf{secrets can enter registers, but not transiently leave them}.
\DefenseName transforms Spectre from a problem that cannot be solved purely in software~\cite{Mcilroy2019}, to a problem that is not easy to solve, but solvable in software. 
For this, \DefenseName requires minimal modifications of applications, compilers, operating systems, and the hardware.
\DefenseName offers full protection for secrets in memory and secrets in registers.
We evaluate the security and performance of \DefenseName.
With its principled approach it inherently mitigates the recently found microarchitectural data sampling attacks on small processor buffers.
Even when over-approximating, we observe no performance overhead for unprotected code and data, and an overhead of \SI{71.14}{\percent} for security-critical applications, which is below the overhead of currently recommended state-of-the-art mitigation strategies.
The actual overhead of \DefenseName is below \SI{1}{\percent} for real-world workloads.
\end{abstract}


\maketitle

\section{Introduction}
\label{sec:introduction}
As arbitrary shrinking of process technology and increasing processor clock frequencies is not possible due to physical limitations, performance improvements in modern processors are made by increasing the number of cores or by optimizing the instruction pipeline.
Out-of-order execution and speculative execution are among the biggest contributors to the performance and efficiency of modern processors.
Out-of-order execution allows processing instructions in an order deviating from the one specified in the instruction stream.
To fully utilize out-of-order execution, processors use prediction mechanisms, \eg for branch directions and targets.
This predicted control flow is commonly called speculative execution. 
However, predictions might be wrong, and virtually any instruction can raise a fault, \eg a page fault.
Hence, in this case, already executed instructions have to be unrolled, and their results have to be discarded.
Such instructions are called transient instructions~\cite{Lipp2018meltdown,Kocher2019spectre,Vanbulck2018,Weisse2018foreshadowNG}.

Transient instructions are never committed, \ie they are never visible on the architectural level. 
Until the discovery of transient-execution attacks, \eg Spectre~\cite{Kocher2019spectre}, Meltdown~\cite{Lipp2018meltdown}, and Fore\-sha\-dow~\cite{Vanbulck2018,Weisse2018foreshadowNG}, they were not considered a security problem. 
These attacks exploit transient execution, \ie execution of transient instructions, to leak secrets. 
This is accomplished by accessing secrets in the transient-execution domain and transmitting them via a microarchitectural covert channel to the architectural domain. 

The original Spectre attack~\cite{Kocher2019spectre} used a cache covert channel to transmit data from the transient-execution domain to the architectural domain. 
However, other covert channels can be used, \eg instruction timings~\cite{Kocher2019spectre,Schwarz2018netspectre}, register contention~\cite{Kocher2019spectre}, branch-predictor state~\cite{Evtyushkin2018BranchScope}, or the TLB~\cite{Kiriansky2018speculative,Schwarz2018netspectre}.
For other covert channels~\cite{Xu2011,Wu2012,Wu2014,Guri2015bitwhisper,Evtyushkin2016RNG,Liu2015,Irazoqui2016Cross,Ge2016,Gruss2016Flush,Maurice2017Hello,Pessl2016,Schwarz2017Timers}, it is still unclear whether they can be used.

Several countermeasures have been proposed against transient-execution attacks, often relying on software workarounds. 
However, many countermeasures~\cite{Yan2018InvisiSpec,khasawneh2018safespec,Kiriansky2018dawg,AMDSpecAnalysis,IntelSpecAnalysis} only try to prevent the cache covert channel of the original Spectre paper~\cite{Kocher2019spectre}. 
This includes the officially suggested workaround from Intel and AMD~\cite{IntelSpecAnalysis,AMDSpecAnalysis} to prevent Spectre variant 1 exploitation. 
However, Schwarz~\etal\cite{Schwarz2018netspectre} showed that this is insufficient. 

State-of-the-art countermeasures can be categorized into 3 classes based on how they try to stop leakage~\cite{Miller2018taxonomy,Canella2018}:

\begin{enumerate}
    \item Mitigating or reducing the accuracy of the covert channel communication, \eg eliminating the covert channel or making gadgets unreachable~\cite{Kiriansky2018dawg,khasawneh2018safespec,Yan2018InvisiSpec}.
    \item Aborting or preventing transient execution when accessing secrets~\cite{IntelSpecAnalysis,ARMSpecAnalysis,AMDSpecAnalysis,Microsoft2018Spectre,Intel2018retpoline,Carruth2018Hardening,Oleksenko2018ysnb,Turner2018retpoline}.
    \item Ensuring that secret data is unreachable~\cite{Reis2018siteisolation,Chromium2018SiteIsolation,Gruss2017KASLR}.
\end{enumerate}

In this paper, we introduce a new type of countermeasure. 
Our approach, \DefenseName, precisely prevents secret data from being used in the transient-execution domain without aborting or preventing transient execution. 
Architecturally, the secret data is still reachable.
However, the secret data is not available when executing in the transient-execution domain.
We show that our approach is efficient and still runs non-dependent instructions out-of-order or speculatively. 
Moreover, we show that our approach effectively prevents all Spectre attacks, but also very recent microarchitectural data sampling attacks~\cite{Schwarz2019ZL,Minkin2019,vanSchaik2019,Schwarz2019STL}.

Implementing \DefenseName in CPUs only requires repurposing one page-table entry bit (\eg  one of the currently unused ones) as a \textit{non-transient bit}. 
Instead of the actual value, the CPU uses a dummy value (\eg `0') when accessing a non-transient memory location during transient execution. 
Additionally, to protect register contents as well, we also introduce a \textit{non-transient bit} per register.
Same as for the memory locations, the CPU will use a dummy value during transient execution instead of the actual register content.

We annotate variables that can hold secrets in the source code.
With compiler and linker support, we propagate this information into the binary, resulting in a separate binary section for secrets.
For this section, the operating system sets the memory mapping to \textit{non-transient}.
We split the stack into an unprotected stack and a transient stack. 
The unprotected stack is marked as \textit{non-transient} to be used as temporary memory by the compiler, \eg register spills, and local variables are moved to the transient stack. 
Thus, there is no performance impact for local variables. 
Preventing leakage only requires a developer to identify the assets, \ie secret values, inside an application. 
Obviously, this is much easier than identifying \emph{all code locations} which potentially leak secret values. 

To emulate the minimal hardware adaptions \DefenseName requires, we over-approximate it via \DefenseNameLight,
a software-only solution which partially emulates the behavior using existing features of commodity CPUs. 
\DefenseNameLight relies on the property that values stored in \textit{uncacheable memory} can generally not be used inside the transient-execution domain~\cite{Eclypsium2018smm,Lipp2018meltdown}, except for cases where the value is architecturally in registers, or microarchitecturally in the load buffer, store buffer, or line fill buffer. 
It is an over-approximation of \DefenseName, yet, does not provide complete protection on commodity systems due to leakage from these two buffers.
Thus, while it does not provide the same protection guarantees, it allows obtaining a loose upper bound for the worst-case performance overhead of the hardware solution.
\DefenseName only requires the annotation of secrets inside the program, it can be easily added to any existing C/C++ program to protect secrets from being leaked via transient-execution attacks. 
In contrast to the software approximation \DefenseNameLight, \DefenseName also inherently protects against microarchitectural data sampling attacks~\cite{Schwarz2019ZL,Minkin2019,vanSchaik2019,Schwarz2019STL}, as leakage is, by design, prevented on the register-level and the state of caches and buffers does not matter.

We evaluate the security of \DefenseName on all known Spectre attacks. 
Due to its principled design, \DefenseName prevents the leakage of secret data in all cases.
The overhead is less than \SI{71.14}{\percent}, which is lower than the overhead of the currently recommended and deployed countermeasures~\cite{IntelSpecAnalysis,ARMSpecAnalysis,AMDSpecAnalysis,Microsoft2018Spectre,Reis2018siteisolation,Chromium2018SiteIsolation,Carruth2018Hardening,Larabel2018stibp,Tkachenko2018ibrs_performance}.
To further support the performance analysis, we extended the Bochs emulator with the \textit{non-transient bits} for registers and page tables and extended it with a cache simulator.
With the hardware extension, the overhead of \DefenseName is below \SI{1}{\percent} for most real-world workloads.

Concurrent to our work, NVIDIA patented a closely related to our design~\cite{Boggs2019memory}.
However, they do not provide protection for registers, but only for memory locations.

\paragraph{Contributions}
The contributions of this work are:

\begin{enumerate}
 \item We propose \DefenseName, a hardware-software co-design for considerate transient execution, fully mitigating transient-execution attacks. 
 \item We show that on all levels only minimal changes are necessary. The proposed hardware changes can be partially emulated on commodity hardware. 
 \item We demonstrate that \DefenseName prevents all known Spectre variants, even if they do not rely on the cache for the covert channel. 
 \item We evaluate the performance of \DefenseName and show that the overhead is lower than the overhead of state-of-the-art countermeasures.
\end{enumerate}

\paragraph{Outline} The remainder of this paper is organized as follows.
In \cref{sec:background}, we provide background information.
\Cref{sec:design} presents the design of \DefenseName.
\Cref{sec:implementation} details our approximate proof-of-concept implementation on commodity hardware. 
\Cref{sec:evaluation} provides security and performance evaluations.
\Cref{sec:discussion} discusses the context of our work.
We conclude our work in \cref{sec:conclusion}.

\section{Background}
\label{sec:background}
In this section, we give an overview of transient execution.
We then discuss known transient execution attacks.
We also discuss proposed defenses and their shortcomings.

\subsection{Transient Execution}
To simplify processor design and to allow superscalar processor optimizations, modern processors first decode instructions into simpler micro-operations (\muops)~\cite{Fog2016}.
With these \muops, one optimization is not to execute them in-order as given by the instruction stream but to execute them \textit{out-of-order} as soon as the execution unit and required operands are available.
Even in the case of out-of-order execution, instructions are retired in the order specified by the instruction stream.
This necessitates a buffer, called reorder buffer, where intermediate results from \muops can be stored until they can be retired as intended by the instruction stream.

In general, software is almost never purely linear but contains (conditional) branches.
Without speculative execution, a processor would have to wait until the branch is resolved before execution can be continued, drastically reducing performance.
To increase performance, speculative execution allows a processor to predict the most likely outcome of the branch using various predictors and continue executing along that direction until the branch is resolved.

At runtime, a program has different ways to branch, \eg conditional branches or indirect calls.
Intel provides several structures to predict branches~\cite{Intel_opt}, \eg Branch History Buffer (BHB)~\cite{Bhattacharya2017perf}, Branch Target Buffer (BTB)~\cite{Lee2017BranchShadowing,Evtyushkin2016ASLR}, the Pattern History Table (PHT)~\cite{Fog2016}, and Return Stack Buffer (RSB)~\cite{Fog2016,Maisuradze2018spectre5,Koruyeh2018spectre5}.
On multi-core CPUs, Ge~\etal\cite{Ge2016} have shown that the branch prediction logic is not shared among physical cores, preventing one physical core from influencing the prediction on another.

Speculation is not limited to branches.
Processors can, \eg speculate on the existence of data dependencies~\cite{Horn2018spectre4}.
In the case where the prediction was correct, the instructions in the reorder buffer are retired in-order.
If the prediction was wrong, the results are squashed, and a rollback is performed by flushing the pipeline and the reorder buffer.
During that process, all architectural but no microarchitectural changes are reverted.
Any instruction getting executed out-of-order or speculatively but not architecturally is called a \textit{transient instruction}.
\textit{Transient execution} may have measurable microarchitectural side effects.

\subsection{Transient Execution Attacks \& Defenses}\label{sec:attacks-defenses}
While transient execution does not influence the architectural state, the microarchitectural state can change.
Attacks that exploit these microarchitectural state changes to extract sensitive information are called transient execution attacks.
So-called Spectre-type~\cite{Kocher2019spectre} attacks exploit different prediction mechanisms, while Meltdown-type~\cite{Lipp2018meltdown,Vanbulck2018} attacks exploit transient execution following a CPU exception.

Kocher~\etal\cite{Kocher2019spectre} first introduced two variants of Spectre attacks.
The first exploits the PHT and the BHB such that the processor mispredicts the code path following a conditional branch.
If the transiently executed code loads and leaks the secret, it is called a Spectre gadget.
Kiriansky and Waldspurger~\cite{Kiriansky2018speculative} extended this attack from loads to stores, enabling transient buffer overflows and, thus, extending the number of possible Spectre gadgets.

Variant 2~\cite{Kocher2019spectre} targets indirect branches and poisons the BTB with attacker-chosen destinations, leading to transient execution of the code at this attacker-chosen destination.
An attacker mistrains the processor by performing indirect branches within the attacker's own address space to the address of the chosen address, regardless of what resides at this location.
Chen~\etal\cite{Chen2018SGXpectre} showed that this can also be exploited in SGX.

For a memory load, the processor checks the store buffer for stored values to this memory location.
\mbox{Variant 4}~\cite{Horn2018spectre4}, Speculative Store Bypass, exploits when the processor transiently uses a stale value because it could not find the updated value in the store buffer, \eg due to aliasing.

SpectreRSB~\cite{Koruyeh2018spectre5} and ret2spec~\cite{Maisuradze2018spectre5} are Spectre variants targeting the RSB, a small hardware stack of recent return addresses pushed during recent \textit{call} instructions.
When a \textit{ret} is executed, the top of the RSB is used to predict the return address.
An attacker can force misspeculation in various ways, \eg by overfilling the RSB, or by overwriting the return address on the software stack.

All of the attacks discussed above have three things in common.
First, they all use transient execution to access data that they would not access in normal, considerate execution.
Second, they use this data to influence the microarchitectural state which can be observed using microarchitectural attacks, \eg \FlushReload~\cite{Yarom2014}.
Third, all are executed locally on the victim machine, requiring the attacker to run code on the machine.
Schwarz~\etal\cite{Schwarz2018netspectre} extended the original Spectre attack with a remote component and demonstrated that the microarchitectural state of the AVX2 unit can be used instead of the cache state to leak data.

Meltdown-type attacks exploit deferred handling of exceptions and do not exploit misspeculation but use other techniques to execute instructions transiently.
Between the occurrence of an exception and it being raised, instructions can be executed transiently that access data retrieved by the faulting instructions.
The original Meltdown attack~\cite{Lipp2018meltdown} exploited the deferred pagefault following a user/supervisor bit violation, allowing to leak arbitrary memory.
A variation of this attack allows an attacker to read system registers~\cite{ARMSpecAnalysis,IntelSpecAnalysis}.
Van Bulck~\etal\cite{Vanbulck2018,Weisse2018foreshadowNG} demonstrated that this problem also applies to other page-table bits, namely the present and the reserved bits.
Canella~\etal{Canella2019} analyzed different exception types, based on Intel's~\cite{Intel_vol3} classification of exceptions as \emph{faults}, \emph{traps}, and \emph{aborts}.
They found that all known Meltdown variants so far have exploited faults, but not traps or aborts.

\paragraph{Defenses}
Since the discovery of Spectre, many different defenses have been proposed.
The easiest and most radical solution would be to entirely (or selectively) disable speculation at the cost of a huge decrease in performance~\cite{Kocher2019spectre}.
Intel and AMD proposed a similar solution by using serializing instructions on both outcomes of a branch~\cite{AMDSpecAnalysis,IntelSpecAnalysis}.
Evtyushkin~\etal\cite{Evtyushkin2018BranchScope} proposed to allow a developer to annotate branches that could leak sensitive data, which are then not predicted.
Unfortunately, on Intel CPUs, serializing branches does not prevent microarchitectural effects such as powering up AVX units, or TLB fills~\cite{Schwarz2018netspectre}.

For mitigating the RSB attack vector, Intel proposes RSB stuffing~\cite{Intel2018retpoline}. 
Upon each context switch, the RSB is filled with the address of a benign gadget.

Google Chrome limits the amount of data that can be extracted by introducing \textit{site isolation}~\cite{Chromium2018SiteIsolation}.
Site isolation relies on process isolation, \ie each site is executed in its own process.
Thus, Spectre attacks cannot leak secrets of other sites. 
Speculative Load Hardening~\cite{Carruth2018Hardening} and YSNB~\cite{Oleksenko2018ysnb} are similar proposals, both limiting speculation by introducing data dependencies between the array access and the condition.

SafeSpec~\cite{khasawneh2018safespec} and InvisiSpec~\cite{Yan2018InvisiSpec} introduce additional shadow hardware for speculation.
The results of transient instructions are only made visible to the actual hardware when the processor determined that the prediction was correct.
Both methods require major changes to the hardware.

DAWG~\cite{Kiriansky2018dawg} is another proposal requiring major hardware changes.
The idea is to partition the cache to create protection domains which are disjoint across ways and metadata partitions.
Additionally to hardware changes, the approach requires changes to the replacement policy and cache coherence protocol to incorporate the protection domain.

All local Spectre variants so far use either \FlushReload~\cite{Yarom2014,Kocher2019spectre,Horn2018spectre4,Koruyeh2018spectre5,Maisuradze2018spectre5} or \PrimeProbe~\cite{Osvik2006,Trippel2018MeltdownPrime} to extract information from the covert channel, requiring access to a high-resolution timer.
Thus, a defense mechanism is to reduce the accuracy of timers~\cite{Microsoft2018edge,Pizlo18,Chromium18mitigations,Wagner18firefox} and eliminate methods to construct different timers~\cite{Schwarz2017Timers}.

To mitigate Spectre variant 2, both Intel and AMD extended the ISA with mechanisms to control indirect branches~\cite{AMDspex_whitepaper,IntelMitigations}, namely Indirect Branch Restricted Speculation (IBRS), Single Thread Indirect Branch Prediction (STIBP), and Indirect Branch Predictor Barrier (IBPB).
With IBRS, the processor enters a special mode and predictions cannot be influenced by operations outside of it.
STIBP restricts the sharing of branch prediction mechanisms among hyperthreads.
IBPB allows to flush the BTB.
Future processors implement enhanced IBRS~\cite{Intel2018retpoline}, a hardware mitigation for Spectre variant 2.
With \textit{retpoline}~\cite{Turner2018retpoline}, Google proposes an alternative technique to protect against branch poisoning by ensuring that the return instruction predicts to a benign endless loop through the RSB.

To mitigate Spectre variant 4, Intel provides a microcode update to disable the speculation on the store buffer check~\cite{IntelMitigations}.
The new feature, called Speculative Store Buffer Disable (SSBD), is also supported by AMD~\cite{AMDssbd_whitepaper}.
ARM introduced a new barrier (\texttt{SSBB}) which prevents loads after the barrier from bypassing a store using the same virtual address before the barrier~\cite{ARMSpecAnalysis}.
Future ARM CPUs will feature a configuration control register that prevents the re-ordering of stores and loads.
This feature is called Speculative Store Bypass Safe (SSBS)~\cite{ARMSpecAnalysis}.

So far, all the proposed defense mechanisms against Spectre attacks either require substantial hardware changes or only consider cache-based covert channels.
In the latter case, an attacker can circumvent the defense by using a different covert channel.
This focus on cache covert channels only and the huge decrease in performance caused by state-of-the-art Spectre defenses shows the necessity for development of efficient and effective defenses.

To mitigate Meltdown, Gruss~\etal\cite{Gruss2017KASLR} proposed KAISER, a kernel modification unmapping most of the kernel space while running in user mode~\cite{Gruss2017KASLR}.
The idea of KAISER has been integrated into all major operating systems, \eg in Linux as KPTI~\cite{LWN_kpti}, in Windows as KVA Shadow~\cite{Ionescu2017Twitter}, and in Apple's xnu kernel as double map~\cite{Levin2012}.
With the PCID and ASID support of modern processors, the performance overheads appear acceptable for real-world use cases~\cite{Gregg2018kpti}. 
Additionally, to mitigate Foreshadow~\cite{Vanbulck2018} on SGX enclaves, microcode updates are necessary.
To mitigate Foreshadow-NG~\cite{Weisse2018foreshadowNG}, several further steps need to be implemented for full mitigation.
The kernel must use non-present page-table entries more carefully, \eg not store the swap disk page frame number there for swapped-out pages.
When using EPTs (extended page tables), the hypervisor must make sure that the L1 cache does not contain any secrets when switching into a virtual machine.
Hence, the defenses for Meltdown are complete but expensive.

\subsection{Taint Analysis}\label{sec:taintanalysis}
Taint tracking is used to track data-flow dependencies on a hardware level~\cite{Chow2004,Song2008}, binary-level~\cite{Cheng2006,Schwartz2010}, or source level~\cite{Shankar2001}.
Taint analysis has a wide range of security applications: detecting vulnerabilities, \eg by tracking untrusted user input; malware analysis, \eg analyzing information flows in binaries; test case generation, \eg automatically generating inputs.
This can be either done statically~\cite{Arzt2014,Wang2008Still} or dynamically~\cite{Newsome2005,Qin2006}.

Dynamic taint analysis allows to track the information flow between sources and sinks~\cite{Schwartz2010}.
Any value that depends on data derived from a tainted source, \eg user input, is considered \textit{tainted}.
Values that are not derived from tainted sources are considered \textit{untainted}.
A policy defines how taint flows as the program executes and how new taints are introduced.
Over-approximation can occur when tainting a value that is not derived from a taint source.

Taint tracking has also been proposed on a hardware level~\cite{Venkataramani2008}, yet not in the context of speculative execution.

\section{Design of \DefenseName}\label{sec:design}

In this section, we present the design of \DefenseName, a considerate transient execution technique.

The idea of \DefenseName is to introduce a new type of memory mappings, namely \textit{non-transient} mappings. 
The \textit{non-transient} option indicates that the mapping contains secrets which must not be accessed within the transient-execution domain.
Consequently, \textit{non-transient} values must not be used in transient operations, neither directly nor in a derived form.
Thus, there cannot be any perturbations of the microarchitectural CPU state which might disclose \textit{non-transient} values via side channels.
To track whether a value is \textit{non-transient} and must be protected, registers also track the \textit{non-transient} state.
To ensure not only the original but also derived values are protected, this information is propagated to the results of operations using these values, until the secret is destroyed, \eg by overwriting it.

A processor with \DefenseName mitigates all speculative execution and microarchitectural data sampling attacks, as the processor cannot use \textit{non-transient} registers anymore.
However, code that exposes information architecturally already is considered out of scope, \eg branching based on secrets.

\DefenseName is a multi-level countermeasure which works on the application-, compiler-, operating-system-, and hardware-level.
An application developer annotates secret values in the source code, which the compiler groups inside the binary and marks as secret. 

Besides annotation of secrets it would also be possible to architecturally define groups of secrets, \eg by defining all userspace memory and user input as secret as proposed by Taram~\etal\cite{Taram2019}.
However, this can be very expensive, and consequently, related work is also investigating annotation-based protection mechanisms~\cite{Yu2019data}.

When the operating system loads the binary, memory regions containing the annotated secrets are marked \textit{non-transient}. 
All subsequent tracking of secrets is done by the hardware. 
The operating system only has to be aware of secret register states on interrupts, \eg context switches. 
Other than these minimal changes, there are no additional adaptions required on any level of the software stack.

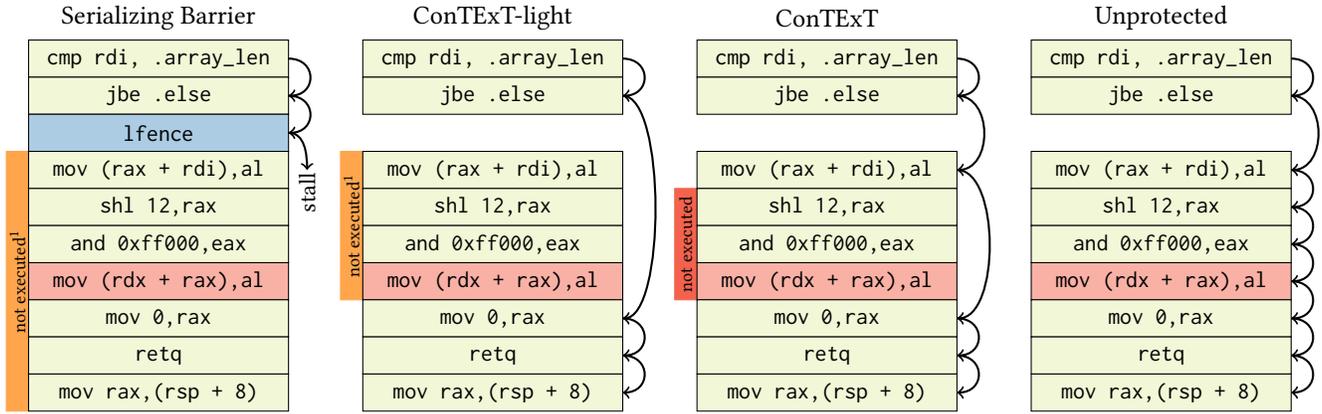
\begin{figure*}[t]
 \centering
 \resizebox{1\hsize}{!}{
\begin{tikzpicture}

\node at (1.75, 6.8) {\large Serializing Barrier};

\draw[fill=orange!70,draw=orange!70] (-0.3,1.5) rectangle (0,5) node[midway,rotate=90] {\footnotesize{not executed$^1$}}; 
\draw[fill=green!20] (0,6) rectangle +(3.5, 0.5) node[pos=.5] {\texttt{cmp rdi, .array\_len}};
\draw[fill=green!20] (0,5.5) rectangle +(3.5, 0.5) node[pos=.5] {\texttt{jbe .else}};
\draw[in=0,out=0,looseness=2,->,thick] (3.5, 6.25) to +(0, -0.5);
\draw[in=0,out=0,looseness=2,->,thick] (3.5, 5.75) to +(0, -0.5);
\draw[fill=blue!40] (0,5) rectangle +(3.5, 0.5) node[pos=.5] {\texttt{lfence}};
\draw[in=90,out=0,->,thick] (3.5, 5.25) to +(0.25, -0.5) node[rotate=90,xshift=-0.3cm] {stall};
\draw[fill=green!20] (0,4.5) rectangle +(3.5, 0.5) node[pos=.5] {\texttt{mov (rax + rdi),al}};
\draw[fill=green!20] (0,4) rectangle +(3.5, 0.5) node[pos=.5] {\texttt{shl 12,rax}};
\draw[fill=green!20] (0,3.5) rectangle +(3.5, 0.5) node[pos=.5] {\texttt{and 0xff000,eax}};
\draw[fill=red!40] (0,3) rectangle +(3.5, 0.5) node[pos=.5] {\texttt{mov (rdx + rax),al}};
\draw[fill=green!20] (0,2.5) rectangle +(3.5, 0.5) node[pos=.5] {\texttt{mov 0,rax}};
\draw[fill=green!20] (0,2) rectangle +(3.5, 0.5) node[pos=.5] {\texttt{retq}};
\draw[fill=green!20] (0,1.5) rectangle +(3.5, 0.5) node[pos=.5] {\texttt{mov rax,(rsp + 8)}};

\begin{scope}[shift={(+4.5,+0)}]

\node at (1.75, 6.8) {\large \DefenseNameLight};

\draw[fill=orange!70,draw=orange!70] (-0.3,3) rectangle (0,5) node[midway,rotate=90] {\footnotesize{not executed$^1$}}; 
\draw[fill=green!20] (0,6) rectangle +(3.5, 0.5) node[pos=.5] {\texttt{cmp rdi, .array\_len}};
\draw[fill=green!20] (0,5.5) rectangle +(3.5, 0.5) node[pos=.5] {\texttt{jbe .else}};
\draw[in=0,out=0,looseness=2,->,thick] (3.5, 6.25) to +(0, -0.5);
\draw[in=0,out=0,looseness=0.5,->,thick] (3.5, 5.75) to +(0, -3);
\draw[in=0,out=0,looseness=2,->,thick] (3.5, 2.75) to +(0, -0.5);
\draw[in=0,out=0,looseness=2,->,thick] (3.5, 2.25) to +(0, -0.5);
\draw[fill=green!20] (0,4.5) rectangle +(3.5, 0.5) node[pos=.5] {\texttt{mov (rax + rdi),al}};
\draw[fill=green!20] (0,4) rectangle +(3.5, 0.5) node[pos=.5] {\texttt{shl 12,rax}};
\draw[fill=green!20] (0,3.5) rectangle +(3.5, 0.5) node[pos=.5] {\texttt{and 0xff000,eax}};
\draw[fill=red!40] (0,3) rectangle +(3.5, 0.5) node[pos=.5] {\texttt{mov (rdx + rax),al}};
\draw[fill=green!20] (0,2.5) rectangle +(3.5, 0.5) node[pos=.5] {\texttt{mov 0,rax}};
\draw[fill=green!20] (0,2) rectangle +(3.5, 0.5) node[pos=.5] {\texttt{retq}};
\draw[fill=green!20] (0,1.5) rectangle +(3.5, 0.5) node[pos=.5] {\texttt{mov rax,(rsp + 8)}};

\end{scope}

\begin{scope}[shift={(+9,+0)}]

\node at (1.75, 6.8) {\large \DefenseName};

\draw[fill=red!80,draw=red!80] (-0.3,3) rectangle (0,4.5) node[midway,rotate=90] {\footnotesize{not executed}}; 
\draw[fill=green!20] (0,6) rectangle +(3.5, 0.5) node[pos=.5] {\texttt{cmp rdi, .array\_len}};
\draw[fill=green!20] (0,5.5) rectangle +(3.5, 0.5) node[pos=.5] {\texttt{jbe .else}};
\draw[in=0,out=0,looseness=2,->,thick] (3.5, 6.25) to +(0, -0.5);
\draw[in=0,out=0,looseness=0.75,->,thick] (3.5, 4.75) to +(0, -2);
\draw[in=0,out=0,looseness=1.25,->,thick] (3.5, 5.75) to +(0, -1);
\draw[in=0,out=0,looseness=2,->,thick] (3.5, 2.75) to +(0, -0.5);
\draw[in=0,out=0,looseness=2,->,thick] (3.5, 2.25) to +(0, -0.5);
\draw[fill=green!20] (0,4.5) rectangle +(3.5, 0.5) node[pos=.5] {\texttt{mov (rax + rdi),al}};
\draw[fill=green!20] (0,4) rectangle +(3.5, 0.5) node[pos=.5] {\texttt{shl 12,rax}};
\draw[fill=green!20] (0,3.5) rectangle +(3.5, 0.5) node[pos=.5] {\texttt{and 0xff000,eax}};
\draw[fill=red!40] (0,3) rectangle +(3.5, 0.5) node[pos=.5] {\texttt{mov (rdx + rax),al}};
\draw[fill=green!20] (0,2.5) rectangle +(3.5, 0.5) node[pos=.5] {\texttt{mov 0,rax}};
\draw[fill=green!20] (0,2) rectangle +(3.5, 0.5) node[pos=.5] {\texttt{retq}};
\draw[fill=green!20] (0,1.5) rectangle +(3.5, 0.5) node[pos=.5] {\texttt{mov rax,(rsp + 8)}};

\end{scope}

\begin{scope}[shift={(+13.5,+0)}]

\node at (1.75, 6.8) {\large Unprotected};

\draw[fill=green!20] (0,6) rectangle +(3.5, 0.5) node[pos=.5] {\texttt{cmp rdi, .array\_len}};
\draw[fill=green!20] (0,5.5) rectangle +(3.5, 0.5) node[pos=.5] {\texttt{jbe .else}};
\draw[in=0,out=0,looseness=2,->,thick] (3.5, 6.25) to +(0, -0.5);
\draw[in=0,out=0,looseness=1.25,->,thick] (3.5, 5.75) to +(0, -1);
\draw[in=0,out=0,looseness=2,->,thick] (3.5, 4.75) to +(0, -0.5);
\draw[in=0,out=0,looseness=2,->,thick] (3.5, 4.25) to +(0, -0.5);
\draw[in=0,out=0,looseness=2,->,thick] (3.5, 3.75) to +(0, -0.5);
\draw[in=0,out=0,looseness=2,->,thick] (3.5, 3.25) to +(0, -0.5);
\draw[in=0,out=0,looseness=2,->,thick] (3.5, 2.75) to +(0, -0.5);
\draw[in=0,out=0,looseness=2,->,thick] (3.5, 2.25) to +(0, -0.5);
\draw[fill=green!20] (0,4.5) rectangle +(3.5, 0.5) node[pos=.5] {\texttt{mov (rax + rdi),al}};
\draw[fill=green!20] (0,4) rectangle +(3.5, 0.5) node[pos=.5] {\texttt{shl 12,rax}};
\draw[fill=green!20] (0,3.5) rectangle +(3.5, 0.5) node[pos=.5] {\texttt{and 0xff000,eax}};
\draw[fill=red!40] (0,3) rectangle +(3.5, 0.5) node[pos=.5] {\texttt{mov (rdx + rax),al}};
\draw[fill=green!20] (0,2.5) rectangle +(3.5, 0.5) node[pos=.5] {\texttt{mov 0,rax}};
\draw[fill=green!20] (0,2) rectangle +(3.5, 0.5) node[pos=.5] {\texttt{retq}};
\draw[fill=green!20] (0,1.5) rectangle +(3.5, 0.5) node[pos=.5] {\texttt{mov rax,(rsp + 8)}};

\end{scope}

\end{tikzpicture}
 }
 \caption{Comparison of \DefenseName with the currently recommended solution against the first Spectre attack example~\cite{Kocher2019}. The leaking access, \ie the only line that must not be executed, is highlighted. Serializing barriers and \DefenseNameLight only provide incomplete protection (1), as parts of the instructions may still be executed if the values are only retrieved from registers, load buffer, store buffer, or line fill buffer.}
 \label{fig:comparison-spectre}
\end{figure*}

The full protection \DefenseName requires small hardware changes, which retrofits mechanisms which already exist in today's CPUs, \ie there is no re-design required.
Moreover, the change is fully backwards compatible with existing hardware and software (\ie applications, libraries, and operating systems).
As hardware changes cannot be conducted on commodity CPUs, we evaluate \DefenseName based on \DefenseNameLight, an over-approximation which only requires software changes. 
As illustrated in \cref{fig:comparison-spectre}, \DefenseName is a more considerate variant of transient execution.

\DefenseName protects secrets which are stored in cache and DRAM, \ie attackers cannot access data from memory locations marked as \textit{non-transient} during transient execution, and registers if they have been filled with data from protected cache or DRAM locations or other protected registers.
\DefenseNameLight cannot protect secrets while they are architecturally stored in registers of running threads, or microarchitecturally in the load buffer, store buffer, or in the line fill buffer. 
With \DefenseNameLight, an attacker can still leak data from these microarchitectural structures.
We only use it to obtain a loose upper bound for the performance overheads of \DefenseName.

\DefenseName is a multi-level countermeasure consisting of 3 major components which we describe in this section:
\begin{enumerate}
\item \textit{non-transient} memory mappings (\cf \Cref{sec:design-mapping}),
\item tracking of \textit{non-transient} data (\cf \Cref{sec:design-tracking}), and
\item software (\ie OS, compiler, and application) support for the hardware features (\cf \Cref{sec:design-software}). 
\end{enumerate}

\subsection{Non-Transient Memory Mappings}\label{sec:design-mapping}
We present three possible implementations of \textit{non-transient} memory mappings, \ie memory mappings which indicate that the values cannot be used during transient execution.\footnote{Concurrent to our work, NVIDIA patented a closely related to our design~\cite{Boggs2019memory}. However, they do not provide protection for registers, but only for memory locations.}
All variants allow integrating \DefenseName into the current architecture while maintaining backwards compatibility, \ie if the operating system is not aware of \DefenseName, the changes have no side effects. 
Hence, to implement \DefenseName \emph{only one of the following} variants has to be implemented.

\paragraph{Currently Reserved Page-Table Entry Bit}
There is already sufficient space to store the \textit{non-transient} bit in the page tables of commodity CPUs. 
On Intel 64-bit (IA-32e) systems, each page-table entry has 64 bits, but the defined maximum physical address only has 52 bits.
However, most processors do not support full 52 bits, but only up to 46 bits, which allows working with up to 64\,TB of physical RAM if the hardware supported it.

\begin{figure}[t]
\centering
\resizebox{\hsize}{!}{
 \begin{tikzpicture}[scale=0.6,yscale=1]
        \node at (0,1.25) {\small \texttt{0}};
        \node at (16,1.25) {\small \texttt{15}};
        \node at (0,-3.3) {\small \texttt{48}};
        \node at (16,-3.3) {\small \texttt{63}};
		\draw[fill=black!0] (0,0) rectangle +(1,1) node[midway] {P};
		\draw[fill=black!0] (1,0) rectangle +(1,1) node[midway] {RW};
		\draw[fill=black!0] (2,0) rectangle +(1,1) node[midway] {US};
		\draw[fill=black!0] (3,0) rectangle +(1,1) node[midway] {WT};
		\draw[fill=black!0] (4,0) rectangle +(1,1) node[midway] {UC};
		\draw[fill=black!0] (5,0) rectangle +(1,1) node[midway] {R};
		\draw[fill=black!0] (6,0) rectangle +(1,1) node[midway] {D};
		\draw[fill=black!0] (7,0) rectangle +(1,1) node[midway] {S};
		\draw[fill=black!0] (8,0) rectangle +(1,1) node[midway] {G};
		\draw[fill=black!10] (9,0) rectangle +(3,1) node[midway] {Ignored};
		\draw[fill=black!5] (12,0) rectangle +(4,1);
		\draw[fill=black!5] (16-16,-1) rectangle +(16,1);
		\draw[fill=black!5,draw=none] (12.02,-0.5) rectangle +(3.96,1);
		\draw[fill=black!5] (32-32,-2) rectangle +(14,1);
		\draw[fill=yellow!60] (32-32+14,-2) rectangle +(2,1) node[midway] {Res.};
		\draw[fill=black!5,draw=none] (0.02,-2.5) rectangle +(13.96,2);
		\draw (16-16+7,-1) node {\Large Physical Page Number};
		\draw[fill=yellow!60] (48-48,-3) rectangle +(4,1) node [midway] {Reserved};
		\draw[fill=black!10] (52-48,-3) rectangle +(7,1) node[midway] {Ignored};
		\draw[fill=black!0] (59-48,-3) rectangle +(4,1) node[midway] {Prot. Key};
		\draw[fill=black!0] (63-48,-3) rectangle +(1,1) node[midway] {X};
\end{tikzpicture}
}
\caption{A page-table entry on x86-64 consists of 64 bits which define properties of the virtual-to-physical memory mapping. Besides the already used bits, physical page number, and ignored bits (which can be freely used), there are 6 physical address bits which are currently reserved for future use since hardware is limited to 46-bit physical addresses. Future processors may support longer physical addresses.}
\label{fig:pte}
\end{figure}
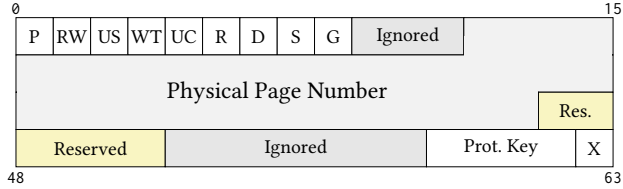

\Cref{fig:pte} shows a page-table entry for x86-64. 
Besides the already used bits, there are the 6 bits between bit 46 and 51, which are currently reserved for future use.
This future use could be the extension of the physical page number if more physical memory is supported in future CPU generations. 
However, it could also be the repurposing of one of the bits (e.g., the last reserved bit) as a \textit{non-transient} bit.
This reduces the theoretical maximum amount of supported memory by factor 2. 
Thus, instead of \SI{4}{\peta\byte}, CPUs could only support \SI{2}{\peta\byte} of physical memory. 
The repurposing of a reserved bit is automatically backwards-compatible, as the reserved bits currently have to be `0'.
Hence, using such a bit does not have any undesirable side effects on legacy software.

\paragraph{Currently Ignored Page-Table Entry Bit and Control Register}\label{sec:design-pte-ign}
An alternative to using one of the reserved bits is to use one of the ignored bits. 
These bits can be freely used by the operating system, thus, simply repurposing them is not possible. 
However, if the feature has to be actively enabled, the operating system is aware of the changed semantics of the specific ignored bit. 
Note that this approach was already taken for several other page-table bits, \eg the protection key and the global bit are enabled via \texttt{CR4} and they are ignored otherwise.
Hence, we also propose enabling the feature using a bit in one of the CPU control registers, \eg \texttt{CR4}, \texttt{EFER}, or \texttt{XCR0}. 
These registers are already used for enabling and disabling security-related features, such as NX (no-execute) or SMAP (supervisor mode access prevention). 
Moreover, these registers still have up to 54 unused control bits which can be used to enable and disable the \textit{non-transient} bit. 

An advantage of repurposing an ignored bit is that CPU vendors do not lose potential address space bits.
That is, this approach is compatible with physical address spaces of up to \SI{4}{\peta\byte} in future hardware. 
However, the approach comes with the limitation that operating systems cannot freely use the retrofitted ignored bit anymore, as it is now used as the \textit{non-transient} bit.

\paragraph{Memory Type using Page-Attribute Table}
A third alternative is to retrofit the Page-Attribute Table (PAT), a processor feature allowing the operating system to reconfigure various attributes for classes of pages. 
The PAT allows specifying the \textit{memory type} of a memory mapping. 
On x86, there are currently 6 different memory types which define the cache policy of the memory mapping. 

\begin{table}
\resizebox{\hsize}{!}{
\renewcommand{\arraystretch}{1}
\begin{tabular}{ccp{0.7\hsize}}
\toprule
\textbf{Value} & \textbf{Type} & \textbf{Description} \\
\midrule
0     & UC   & Strong uncacheable, never cached \\ 
1     & WC   & Write Combining (subsequent writes are combined and written once) \\ 
2     & \textbf{\textit{NS}} & \textbf{\textit{Non-transient, cannot read in transient execution domain}} \\ 
3     & -    & Reserved \\ 
4     & WT   & Write Through (reads cached, writes written to cache and memory) \\ 
5     & WP   & Write Protected (only reads are cached) \\ 
6     & WB   & Write Back (reads/writes are cached) \\ 
7     & UC-  & Uncacheable, overwritten by MTRR \\
\bottomrule
\end{tabular}
}
\caption{The currently supported memory types which can be used in the PAT (Intel 64-bit systems), and the additional non-transient type (bold-italic) as new memory type.}
\label{tab:pat}
\end{table}

\Cref{tab:pat} shows the memory types which can be set using the PAT, including our newly proposed non-transient memory type. 
The PAT itself provides 8 entries for memory types. 
Such a PAT entry is applied to a memory mapping via the 3 page-table-entry bits `3' (write through), `4' (uncacheable), and `7' (PAT). 
These 3 bits combined to a 3-bit number select one of the 8 entries of the PAT. 

Thus, to apply the non-transient memory type to a memory mapping, the OS sets one of the PAT entries to the non-transient memory type `2'. 
Then, this PAT entry can be applied through the existing page-table bits to any memory mapping. 
As the PAT supports 8 entries, and there are currently only 6 memory types (7 if the non-transient type is included), it is still possible to use all supported memory types concurrently on different pages, \ie the approach is fully backwards-compatible. 

An advantage of this approach is that no semantic changes have to be made to page-table entries, \ie all bits in a page-table entry keep their current meaning.
However, this variant may require more changes in the operating system, as \eg Linux already utilizes all of the PAT entries (some memory types are defined twice). 

\subsection{Secret Tracking}\label{sec:design-tracking}

\textit{Non-transient} mappings ensure that \textit{non-transient} memory locations cannot be accessed during transient execution. 
However, we still need to protect secret data that is already loaded into a register.
Registers in commodity CPUs do not have a memory type or protection.
Thus, we require changes to the hardware to implement protection of registers. 
Based on patents from Intel~\cite{Intel2014TaintTracking}, VMWare~\cite{VMWare2013TaintTracking}, and NVIDIA~\cite{Boggs2019memory}, we expect such tracking features to be implemented in future CPUs.
Venkataramani~\etal\cite{Venkataramani2008} proposed a technique that also taints registers, however, to mitigate architecturally and functionally correct behavior rather than overly eager speculative execution.

\paragraph{Tainting Registers}
For \DefenseName, we introduce one additional \textit{non-transient} bit per register, \ie a \textit{taint} (\cf \cref{sec:taintanalysis}).
The \textit{non-transient} bit indicates whether the value stored in the register is \textit{non-transient} or not. 
A register is either entirely \textit{non-transient} or entirely not at all.
The taint generally propagates from memory to registers and from registers to registers.
The rationale behind this is that results of operations on secret data have to be considered secret as well.
Accessing only parts of a tainted register, \eg \texttt{eax} instead of \texttt{rax}, still copies the taint from the source register to the target register and taints the entire target register, as we only have a single \textit{non-transient} bit per register.
This is also true for taint propagation in any other use of a tainted register.

We keep taint propagation very simple and consider only instructions with registers as destination operands.
If any \textit{non-transient} memory location is used as a source operand to an instruction, the instruction taints the destination registers, \ie the \textit{non-transient} bit is set for every destination register.
Similarly, if any \textit{non-transient} register is used as a source operand to an instruction, the instruction also taints the destination registers.
Thus, if a secret is loaded into a register, it is tracked through all register operations. 

The taint is not propagated if the destination operand(s) are memory location(s), as all memory locations already have a \textit{non-transient} bit managed by the operating system.

\paragraph{Untainting Registers}
There are not only operations which taint registers, but also operations which untaint registers. 
Replacing the entire content of a register without using \textit{non-transient} memory or registers, untaints the register. 
We do this to avoid over-tainting registers, a problem pointed out in earlier works~\cite{Slowinska2009}.
In particular, all immediate or untainted values which replace the content of a register, untaint the register. 
Writing a tainted register to a normal memory location, \ie a memory location which is not marked as \textit{non-transient}, also untaints the register. 
The rationale behind this is that if registers are spilled to normal (\ie insecure) memory locations, a potential secret can be leaked anyway. 
If such a memory operation happens unintentionally, it is a bug in the program and has to be fixed at the software level. 
In many cases, however, this will be intentional behavior, as the programmer decided that the register does not contain a secret anymore.
For instance, the output of a cryptographic cipher does not need protection from transient execution attacks. 
Thus, the automated untainting keeps the number of tainted registers small. 

\paragraph{Taint Propagation across Memory Operations}
As the taint bit is an additional bit for each register, it can only be propagated to other registers, not to memory.
If an operation writes a secret (\ie tainted) register to the memory, the taint bit is irrecoverably lost. 
While this is intended if the developer explicitly writes values to memory, it might have undesirable consequences if this happens implicitly, \eg due to the inner workings of the compiler. 
In \Cref{sec:design-software}, we introduce the required changes to the compiler which ensure that non-transient values are never spilled to transient memory locations accidentally. 

However, the compiler inevitably still has to temporarily store (insecure) registers within memory regions marked as \textit{non-transient}. 
With the solution as described so far, we would over-approximate and taint more and more registers over time by spilling them to \textit{non-transient} memory locations and reading them back from there.
Hence, spilling registers is not a security problem (\ie tainted registers are never untainted, only untainted registers are tainted), but a loss in performance due to unnecessarily tainted registers. 

\paragraph{Optimizing Performance via Caching}\label{sec:opt_perf_caching}
To prevent this potential performance loss, we propose an additional change to the cache to reduce the impact of the taint over-approximation. 
We introduce one additional bit per 64 bits to the cache, \ie 8 additional bits per \SI{64}{\byte} cache line. 
This allows us to store the register-taint information transparently in the cache. 
Whenever a register is written to \textit{non-transient} memory, the taint bit of the register is stored in the corresponding cache line. 
When reading from memory, the bit stored in the cache line has precedence over the information from the TLB, \ie the cache overwrites the taint bit defined by the memory mapping. 
The information in the cache allows the hardware to temporarily keep track of the taint information of a register if the register value is moved to the stack. 
This happens, \eg if register values are spilled on the stack, exchanged via the stack, or upon function calls. 

Evicting the cache line corresponding to a register is never a security issue. 
An evicted cache line only loses the information that a register was \textit{not tainted}. 
Thus, if the cache line is evicted, the registers become automatically tainted. 

\subsubsection{Taint Control}
Besides the automated tainting and untainting of registers, \DefenseName provides a privileged interface to modify the taint of registers. 
This interface is necessary for the operating system to save and restore taint values upon context switches. 

A straightforward solution would be to introduce new instructions in the ISA. 
However, we try to keep the hardware changes to a minimum, especially changes which are not hidden in the microarchitecture.  
Hence, we propose instead to use model-specific registers (MSR) to access the taint information of registers. 

\paragraph{Read/Write Taint}
To read and write the current taint information of all registers, we introduce an MSR \texttt{IA32\_TAINT}. 
The taint bit of every register directly maps to one bit of this MSR, which allows the operating system to read and write all taint bits in a single operation. 
As there are only 56 registers (16 general purpose, 8 floating point, 32 vector) which have to be tracked, one 64-bit MSR is sufficient to read or write all taint bits at once. 

\paragraph{Interrupt Handling}\label{sec:int_handling}
MSRs can only be accessed indirectly, using an instruction (\ie \texttt{rdmsr} on x86), and require registers both to specify the MSR and as source and destination operands.
On an interrupt, the first thing to save should be the \texttt{IA32\_TAINT} MSR, because it contains the taints of the previous context.
However, as registers must not be clobbered in the interrupt routine, all the registers used in the interrupt handler have to be saved first. 
We resolve this problem by automatically copying the \texttt{IA32\_TAINT} to an additional MSR, \texttt{IA32\_SHADOW\_TAINT}, on every interrupt. 
This ensures that the taint of all registers is preserved before any taint is potentially modified by a register operation in the interrupt handler. 
The \texttt{IA32\_SHADOW\_TAINT} can then be treated like any other register, \eg the operating system can save it into a kernel structure upon a context switch. 

When returning from an interrupt, the CPU restores the values from \texttt{IA32\_SHADOW\_TAINT} to the register taint values. 
Hence, with this mechanism, we ensure that an interrupt does not influence the taint value of any register. 
This also works for the unlikely event of nested interrupts, \ie if an interrupt is interrupted by a different interrupt.
The only critical region in such a case is if the first interrupt has not yet locally saved the \texttt{IA32\_SHADOW\_TAINT} MSR, and the second interrupt overwrites the MSR. 
However, as long as within this critical region (\ie the time window between first interrupt and second interrupt) no register is untainted, there can be no leakage. 
In \Cref{sec:design-software-os}, we show that this situation can be avoided solely in software.

\subsection{Software Support}\label{sec:design-software}
We propose changes to applications, compilers, and operating systems to leverage the hardware extensions introduced in \Cref{sec:design-mapping} and \Cref{sec:design-tracking}. 
The idea is that application developers annotate secrets in their applications. 
The annotations are processed by the compiler and then forwarded to the operating system to establish the correct memory mappings (\cf \Cref{sec:design-mapping}). 

\paragraph{Compiler}
The compiler parses the annotations of secrets. 
The secrets identified this way are allocated inside a dedicated section of the binary. 
The compiler marks this section as \textit{non-transient}.
The operating system maps this section from the binary using a \textit{non-transient} memory mapping.

Besides parsing the annotations, our modified compiler ensures that it never spills data from registers marked as secret into unprotected memory. 
Otherwise, an attacker could leak the spilled secrets from memory.
Still, it is unavoidable that the compiler spills registers to memory, \eg to preserve register contents over function calls. 
Furthermore, due to the calling convention, some (possibly secret) values have to be passed over the stack. 
Hence, we have to assume that the stack contains secrets. 
As a consequence, the stack has to be mapped using a \textit{non-transient} memory mapping as well. 

To reduce the performance impact of a \textit{non-transient} stack, we modify the compiler to only use the \textit{non-transient} stack if really necessary. 
This \textit{non-transient} stack only contains register spills, possibly function arguments, and return values. 
All other values are stored at a different memory location, the \textit{unprotected} stack. 
This concept is similar to the \safestack~\cite{Kuznetsov2014CPI} and our implementation even reuses parts of the \safestack infrastructure of modern compilers. 
The difference to \safestack, where only ``unsafe'' memory allocations (\eg buffers) are stored on the \safestack, is that we move all variables normally allocated on the stack to the \textit{unprotected} stack. 
Thus, for \DefenseName, only the absolute minimum is stored on the \textit{non-transient} stack, \eg return addresses.
By only moving local variables to the \textit{unprotected} stack, and leaving return addresses and function arguments on the stack, we do not break ABI compatibility with existing binaries. 
Thus, a developer can still use external libraries without recompiling them, and libraries compiled for \DefenseName can be used in ordinary unprotected applications. 

Moving local variables from the stack to a different memory location does not impact the runtime of the application and even gives additional protection against memory-corruption attacks~\cite{Kuznetsov2014CPI}.

\paragraph{Operating System}\label{sec:design-software-os}
For \DefenseName, the operating system is in charge of setting up \textit{non-transient} memory mappings. 
As the operating system parses the binary, it can directly set up the \textit{non-transient} memory mappings which are marked as such by the compiler. 
The operating system requires additional small changes. 
The operating system has to save and restore taint values on context switches. 
The hardware already saves the current taint value of all registers into the \texttt{IA32\_SHADOW\_TAINT} MSR upon interrupts. 
Thus, the operating system only has to read this register and save it together with all other saved registers. 

\begin{listing}[t]
\begin{lstlisting}[language={[x64]Assembler},label={lst:iret}]
 pushall
 rep xor rcx, rcx ; clear rcx, rep prefix keeps taint
 add rcx, IA32_TAINT
 rdmsr ; taint in rax, rdx
 [...]
 popall
 push rax, rcx, rdx
 mov rcx, IA32_TAINT ; also updates IA32_SHADOW_TAINT
 wrmsr ; old taint in rax, rdx
 pop rax, rcx, rdx
 iret ; restores IA32_SHADOW_TAINT to registers
\end{lstlisting}
\caption{(Pseudo-)assembly for saving and restoring the taint MSR without destroying the taint of any other register during a context switch.}
\end{listing}

As interrupts can be interrupted by other interrupts, \eg a normal interrupt can be interrupted by a non-maskable interrupt (NMI), there is a critical section between reading the MSR and saving the result. 
If registers are untainted in this section, a nested interrupt would lose the taint information as it overwrites the \texttt{IA32\_SHADOW\_TAINT} MSR. 
However, if registers are not untainted in this section, no taint information can be lost. 
Hence, we have to initialize the registers required to read the MSR in a way which does not destroy the taint. 
For this purpose, we define that the \texttt{rep} prefix for arithmetic and logical operations on registers, preserves the taint.
\Cref{lst:iret} shows (pseudo-)assembly code which prepares the registers with the required immediate values. 
Generally, overwriting a register with an immediate or by using an idiom, \eg \eg \texttt{xor rax,rax}, generally untaints the register.
However, the \texttt{rep} prefix prevents the untainting here.

In addition to the context switch, the operating system has to flush the cache when the content of a \textit{non-transient} memory is initially loaded from the binary. 
This is important as the initial data transfer to the memory page is not done through the \textit{non-transient} user-space mapping.
Thus, the operating system has to either disable the cache before this operation or flush the corresponding cache lines afterwards. 
This functionality is already present in the x86 ISA and supported by modern operating systems, thus, there is no further change required.

\section{Implementation of \DefenseName}\label{sec:implementation}

In this section, we present our implementation of both \DefenseName and \DefenseNameLight, which we use for the evaluation (\cf \Cref{sec:evaluation}). 
As we cannot change real x86 hardware or emulate the hardware changes required for \DefenseName on commodity hardware, we opted for a hardware simulation of our changes using a full-system emulator (\cf \Cref{sec:impl-hw}). 
While this does not allow to measure performance by measuring the runtime, it allows measuring performance in the number of memory accesses, \textit{non-transient} memory accesses, taint over-approximations, etc., for real-world benchmarks.

For \DefenseNameLight, we present a method to partially emulate the non-transient memory mapping behavior on commodity hardware by retrofitting uncacheable memory mappings. 
Thus, in \Cref{sec:impl-sw}, we present an open-source proof-of-concept implementation of \DefenseNameLight which can already be used and evaluated on commodity hardware.
However, \DefenseNameLight does not provide complete security guarantees, as secrets can still be leaked from registers, line fill buffers, load buffers, and store buffers.
\DefenseName inherently protects also against the recent attacks on microarchitectural buffers~\cite{Schwarz2019ZL,Minkin2019,vanSchaik2019,Schwarz2019STL}, as \DefenseName prevents leakage from registers---the state of microarchitectural buffers does not matter.

\subsection{Hardware Simulation}\label{sec:impl-hw}

We simulated \DefenseName using the open-source x86-64 emulator Bochs~\cite{Lawton1996bochs} to get as close as possible to functionally extending a real x86-64 processor with our features; \textit{non-transient} memory mappings (\cf \Cref{sec:design-mapping}) as well as secret tracking (\cf \Cref{sec:design-tracking}). 
We incorporated hardware and behavioral changes in our \DefenseName-enabled Bochs.

\paragraph{Hardware Changes}
To support secret tracking, a few minor hardware changes are required.
Mostly, these are single bits to track whether a register is \texttt{non-transient}. 
These bits are required in every page-table entry, TLB entry, and register.
Furthermore, we introduce additional bits per cache line to minimize the performance cost of register spills (\cf \Cref{sec:opt_perf_caching}).

\textit{Page-Table Entry.}
To distinguish \textit{non-transient} from normal memory mappings, we have to mark every memory mapping accordingly in the PTE. 
For backwards- and future-compatibility, repurposing one of the ignored bits is the best choice (\cf \Cref{sec:design-pte-ign}). 
If this bit is set, we treat the memory mapping as a region which may contain secrets. 

\textit{Translation Look-aside buffer.}
For performance reasons, modern CPUs cache page-table entries in the TLB. 
Consequently, we need an additional \textit{non-transient} bit in the TLB, caching the bit of the page-table entry.
In Bochs, caching of page-table entries is also implemented as a TLB-like structure allowing the simulated hardware to automatically transfer the added bit from the PTE to the TLB.
Thus, for cached page-table entries, memory accesses use the cached \textit{non-transient} bit from the TLB.

\textit{Cache.}
Bochs only implements an instruction cache, but no data cache which plays a vital role in our design to cache taint information (\cf \Cref{sec:design-tracking}).
Hence, we extended Bochs with data cache emulation by implementing an 8-way (inclusive) last-level cache.
As the exact eviction strategy is unknown~\cite{Gruss2016Row}, we used LRU as a good approximation as it has been used in Intel CPUs until Ivy Bridge~\cite{Gruss2016Row}. 
In our emulated cache, we added 8 taint bits per cache line. 

\textit{Model-Specific Registers.}
As described in \Cref{sec:design-tracking}, we added two new MSRs to Bochs.
Accesses to \texttt{IA32\_TAINT} are directly mapped to the taint bits of the registers, allowing the operating system to read and write all at once.
To save the current taint state on interrupts (\Cref{sec:int_handling}), we ensure data consistency between the two MSRs; a write to \texttt{IA32\_TAINT} also (atomically) updates \texttt{IA32\_SHADOW\_TAINT}. 
This enables us to implement secure context switches (\cf \Cref{lst:iret}).

\paragraph{Behavioral Changes}
All behavioral changes are only enabled if the operating system supports and enables \DefenseName using the corresponding bit in the control register (\cf \Cref{sec:design-mapping}). 
However, taint tracking is enabled unconditionally as it happens implicitly without additional cost. 
This applies to all operations which transfer data from memory to registers or from registers to registers. 
In our proof-of-concept implementation, we added the taint tracking to \SIx{368} out of \SIx{557} instructions implemented in Bochs. 
If no memory mapping is marked as \textit{non-transient}, then no register can be tainted.
Thus, taint tracking simply has no effect if there is no operating system support. 

\subsection{\DefenseNameLight}\label{sec:impl-sw}
In addition to the hardware emulation for \DefenseName, we implemented \DefenseNameLight (\cf \Cref{sec:design}) for Linux.
Our implementation of \DefenseNameLight consists of two parts, a kernel module and a runtime library.
For the full \DefenseName, we provide a compiler extension that minimizes performance penalties of register spills.

For the proof of concept, we emulate \textit{non-transient} memory mappings via \textit{uncacheable} memory mappings.
Uncacheable memory can generally not be accessed inside the transient execution domain~\cite{Eclypsium2018smm,Lipp2018meltdown}. 
Lipp~\etal\cite{Lipp2018meltdown} observed the only exception where memory despite being marked as uncacheable can be read during transient execution:
In the case that an attacker can issue a legitimate load of the target address in parallel on another hyperthread running on the same physical core as the attack, the memory content still can be leaked.
However, opposed to \DefenseName, \DefenseNameLight does not protect secrets while they are architecturally stored in registers of running threads.
Thus, the security guarantees of \DefenseNameLight still hold in this case.

\paragraph{Kernel Module}
We opted to implement the operating-system changes as a kernel module for compatibility with a wide range of kernels. 
The kernel module is responsible for setting up \textit{non-transient} memory mappings. 
As our proof-of-concept implementation relies on uncacheable memory, we do not retrofit page-table bits but use the page-attribute table to declare a memory mapping as uncacheable. 

The kernel module provides an interface for the runtime library (\cf \Cref{sec:impl-sw-rt}) to set up \textit{non-transient} memory mappings. 
This allows keeping the changes in the kernel space minimal as most of the logic and parsing can be implemented in user space. 
The kernel module ensures that the page-attribute table contains an \textit{uncacheable} (UC) entry by reprogramming the page-attribute table if this is not already the case. 
If the runtime library requests a mapping to be marked \textit{non-transient} via the kernel-module interface, the page-table entry is modified to reference the UC entry in the page-attribute table. 
Subsequently, the corresponding TLB entry is flushed. 
We do not flush all cache lines of the mapping, as this would incur additional overhead. 
Thus, the developer (or runtime library) has to take care that values stored on pages marked as \textit{non-transient} are not cached before they are marked as \textit{non-transient}. 

\paragraph{Runtime Library}\label{sec:impl-sw-rt}
The runtime library sets up all static and dynamic \textit{non-transient} memory mappings via the kernel-module interface. 
Our proof-of-concept runtime library supports C and C++ applications and can be included as a single header file. 
The header file provides a keyword, \texttt{nospec}, to annotate variables as secrets. 
This keyword ensures that the linker allocates the variables in a dedicated \texttt{secret} section in the ELF binary. 
Moreover, the header file registers a constructor function which is executed before the actual application, to initialize \DefenseName at runtime. 

When the application starts, the runtime library identifies all memory mappings in the \texttt{secret} section from the ELF binary. 
These memory mappings are then set to \textit{non-transient} (\ie uncacheable) using the kernel module.

The runtime library is only active on application startup and does not influence the application during runtime. 
During runtime, it is only used if the developer requests dynamic \textit{non-transient} memory. 
For this purpose, the runtime library provides a \texttt{malloc\_secure} and \texttt{free\_secure} function. 
These functions mark the allocated memory immediately as \textit{non-transient}.

\paragraph{Compiler}
For the full \DefenseName with hardware support, we also need compiler support.
We extended the LLVM compiler~\cite{DBLP:conf/cgo/LattnerA04} in version 8.0.0 to not use the stack for local variables, but move them to a different part of the memory which we refer to as \textit{unprotected stack}. 
The normal stack is marked as \textit{non-transient} to not leak temporary variables and function parameters the compiler puts on the stack. 
Thus, to reduce the performance impact, we allocate local variables which are defined by the developer in the unprotected stack, which is not marked as \textit{non-transient}. 

Our implementation is based on the already existing \safestack extension~\cite{Kuznetsov2014CPI}. 
We modify the heuristics to not move only specific but all user-defined variables from the \textit{non-transient} stack to the unprotected stack (\safestack in the original extension). 
Allocations coming from function parameters and registers spills are put on the \textit{non-transient} stack.

\section{Evaluation}\label{sec:evaluation}
In this section, we evaluate \DefenseName and \DefenseNameLight with respect to their security properties and their performance. 
We evaluate \DefenseName on our modified Bochs emulator, and \DefenseNameLight on a Lenovo T480s (Intel Core i7-8650U, \SI{24}{\giga\byte} DRAM) running Ubuntu 18.04.1 with kernel version 4.15.0. 

\subsection{Security}
We generally assume that the operating system is trusted, as it handles the \textit{non-transient} memory mappings.
First, we explain how \DefenseName can be used to protect against all transient-execution attacks, and how current commodity hardware can be retrofitted to partially emulate \DefenseName.
Second, we show the limitations of \DefenseName.

\subsubsection{Security of \DefenseName}
The security guarantees of \DefenseName are built on two assumptions: the application developer correctly annotated all secrets as such, and the application does not actively leak secrets (\eg by writing them to memory locations not marked as \textit{non-transient}). 
For the evaluation, we distinguish two cases, based on whether the secret values are used architecturally in the application or not while an attacker mounts a transient-execution attack. 

\paragraph{Architecturally Unused Secrets}
A secret is architecturally unused if the secret is only stored in a \textit{non-transient} memory region, \ie there is no part of the secret which is stored in a register, cache, or normal memory region. 
For example, this is the case if the secret was not used by the time of an attack. 
However, the application can also be in such a state although the secret has already been used in the past. 
If all traces of the secret in normal memory or the cache are already overwritten (or evicted), the application returns again to the state where secrets are architecturally unused. 

In this state, an attacker can only target the secret itself and not an unprotected copy of it. 
It is clear that such an attack cannot be successful, as---per-definition---transiently executed code cannot retrieve the value from a \textit{non-transient} memory region.
Hence, \DefenseName is secure, if its implementation fulfills this property.

\paragraph{Architecturally Used Secrets}
If the entire secret, or parts of it, are stored in a register, cache, or a memory region not marked as \textit{non-transient}, the secret is considered architecturally used.
In this case, an attacker can target any unprotected copy of the secret, not only the original secret stored in the \textit{non-transient} memory region. 
However, an attack fails if the target is marked as secret, \eg by a \textit{non-transient} memory mapping, tainted register, or tainted cache line. 

If a \textit{non-transient} memory region is loaded into a register, the register is tainted and, thus, it cannot be targeted. 
Moreover, the taint is also applied to the corresponding cache line and TLB entry. 
Any register-to-register operation which copies the secret, also copies the taint. 
Similarly, an operation which copies the secret to a \textit{non-transient} memory region is also secure. 
Such operations include, for example, register spills to the stack, temporary storage of registers in local variables, or secrets as function arguments (depending on the calling convention). 
Tainted registers can only be untainted by destroying their content, \ie overwriting them with non-secret values. 
Overwriting a register with an immediate or by using an idiom, \eg \eg \texttt{xor rax,rax}, generally untaints the register.
Using the \texttt{rep} prefix on arithmetic or logical register operations preserves the taint.

Thus, registers can not be untainted while containing a secret.
However, over-approximation can lead to more tainted registers than necessary. 

Operations which copy the secret to a memory region not marked as \textit{non-transient} could be attacked. 
However, such operations are never implicitly generated by the compiler, as the compiler only uses the stack as temporary memory. 
Thus, such an operation has to be explicitly defined by the application developer, which violates the assumption that the application does not actively leak secrets. 

A remaining scenario is the context switch of the application with used secrets. 
In such a case, the application is stopped by the operating system, and the current register content is saved to the kernel. 
As the operating system is aware of register taints, and also considered trusted, it can leverage the taint saving mechanism described in \Cref{sec:int_handling}. 
The registers can again be saved in a \textit{non-transient} memory region to prevent transient-execution attacks on the saved registers.
When returning from the kernel, all registers are first tainted (an over-approximation, as they are restored from a \textit{non-transient} stack), but the original taint is restored just before the end of the context switch. 
Thus, registers containing secrets are always tainted and cannot be targeted.

\subsubsection{Security Limitations of \DefenseNameLight}\label{sec:eval-impl-sec}
As \DefenseNameLight is implemented using uncacheable memory, we evaluated the security properties of uncacheable memory regarding transient execution. 
In our experimental setup, we mark a memory mapping as uncacheable using the PAT. 
Using \FlushReload, we verified that the memory mapping is actually uncacheable. 

Unfortunately, even with uncacheable mappings, secrets can still be leaked if the data is already in a register.
Similarly, if the data is currently in the load buffer or store buffer, because of other operations on the processor (\eg prefetching, speculative execution, architectural accesses), the data can still be leaked.
If the data is in the line fill buffer, again because of other operations on the processor, it can also be accessed like in a regular Spectre or Meltdown attack.
Consequently, \DefenseNameLight does not provide the same security guarantees as \DefenseName.
\DefenseName inherently protects against the microarchitectural data sampling attacks~\cite{Schwarz2019ZL,Minkin2019,vanSchaik2019,Schwarz2019STL} by preventing leakage from registers.
The opportunistic and incorrect loading of registers from microarchitectural buffers is thus not a security problem anymore.

\subsubsection{Limitations}\label{sec:limitations}
\DefenseName can only be effective if used correctly by the application developer, \ie if the developer marks all secrets as secret and does not actively leak secrets. 
However, even if used correctly, there are certain limitations which mostly result from a trade-off between performance and security. 
In the following paragraphs, we point out where application developers must take care to not accidentally leak secrets.

\DefenseName does not support tainting registers which are used to steer the control flow, \eg the instruction pointer or the flags register. 
Hence, the application developer should be careful to not introduce higher order leakage through these registers.
This is a sound reasoning because: If the control flow depends on the secret, the code is inherently not side-channel resilient, \ie other side channels such as cache attacks can already be used to extract the secret. 

Instructions such as CRC32 might also leak secrets if a secret value is used as input either directly, or in combination with an attacker-known value. 
However, as this is again a secret-dependent operation, the developer has to ensure that this does not leak any secrets. 

Another responsibility of the developer is that secret values are not actively copied to memory locations not marked as \textit{non-transient}. 
This cannot be prevented by either the compiler or the hardware, as it is often necessary, \eg the tainted output of a crypto operation (ciphertext) is not secret anymore and can be written to normal memory.

\paragraph{\DefenseNameLight}
As \DefenseNameLight is only a partial emulation of \DefenseName, it comes with some limitations compared to \DefenseName. 
The largest difference to \DefenseName is that secrets in registers, the load buffer, the store buffer, and the line fill buffer are not protected. 
Thus, if a secret is in one of these microarchitectural structures, it remains susceptible to transient-execution attacks.

\subsection{Performance}\label{sec:eval-performance}
We evaluated the performance of \DefenseNameLight as a loose upper bound for the performance overhead of \DefenseName.
We also evaluate the performance overhead of \DefenseName based on our full-system emulation in Bochs.
The SPECspeed 2017 evaluation for the \textit{unprotected stack} of \DefenseName is performed on an i7-8700K machine and all other evaluations are performed on an i7-8650U machine. Both systems run Ubuntu Linux 18.04.1 with kernel 4.15.0.

We evaluated the software implications of our proposed hardware changes using our modified version of Bochs and a modified Linux kernel, based on kernel version 4.15.
For the Linux kernel, we only had to modify 52 lines in 9 files to support the save and restore of register taints on context switches. 
These small changes result in a negligible performance overhead on context switches, \eg for syscalls. 

The latency of syscalls increases by a constant value, which is \SIx{48} cycles (averaged over \SI{500000} syscall invocations).
On a standard Ubuntu Linux installation we observed between \SI{3000} and \SI{5000} syscalls per second on average while performing regular office tasks.
On our test system, we observe an overhead on the system load of around \SI{0.01}{\percent} at this syscall rate.
The highest syscall rates observed for real-world use cases at Netflix, was reported to be around \SI{50000} syscalls per second~\cite{Gregg2018kpti}.
On our test system, we observe an overhead on the system load of around \SI{0.13}{\percent} at this syscall rate.

\paragraph{Compiler Extension}
We evaluated the impact of the \textit{unprotected stack} of \DefenseName using the SPECspeed 2017 integer benchmark~\cite{SpecCPU2017}. 
\Cref{tab:speccpu_results} shows that similarly to the original \safestack implementation~\cite{Kuznetsov2014CPI}, the resulting performance overhead is \SI{1.26}{\percent} on average and in the worst case \SI{5.13}{\percent}. 
\begin{table}[t]
\resizebox{\hsize}{!}{
\renewcommand{\arraystretch}{1}
\scriptsize{
  \begin{tabular}{l rr r}
    \toprule
    \multirow{2}{*}{\textbf{Benchmark}} & \multicolumn{2}{c}{\textbf{SPEC Score}} & \textbf{Overhead} \\
    & Baseline & \DefenseName  & [\%]     \\
    \midrule
    600.perlbench\_s  & $7.03$  & $6.86$  & $+2.42$ \\
    602.gcc\_s        & $11.90$ & $11.80$ & $+0.84$    \\
    605.mcf\_s        & $9.06$  & $9.16$  & $-1.10$ \\
    620.omnetpp\_s    & $5.07$  & $4.81$  & $+5.13$ \\
    623.xalancbmk\_s  & $6.06$  & $5.95$  & $+1.82$ \\
    625.x264\_s       & $9.25$  & $9.25$  & $ 0.00$ \\
    631.deepsjeng\_s  & $5.26$  & $5.22$  & $+0.76$ \\
    641.leela\_s      & $4.71$  & $4.64$  & $+1.48$ \\
    648.exchange2\_s 	& \multicolumn{3}{c}{\textit{would require Fortran runtime}} \\
    657.xz\_s         & $12.10$ & $12.10$ & $ 0.00$ \\
    \midrule
    \textbf{Average}  &       &       & $+1.26$\\
    \bottomrule
  \end{tabular}
}
}
  \caption{Performance evaluation of the \textit{unprotected stack} of \DefenseName using the SPECspeed 2017 integer benchmark. The baseline was compiled with the unmodified compiler, the \DefenseName run uses our modified LLVM compiler.}
  \label{tab:speccpu_results}
\end{table}

These results are not surprising, as only addresses of variables change.
This only requires very little runtime code for maintaining a second stack pointer.
Thus, the small performance overhead is mostly due to the setup time for the additional \textit{non-transient} stack. 

We furthermore evaluated the performance impact introduced by the \textit{non-transient} stack. 
As a baseline we consider the case where we only have one \textit{non-transient} stack and compare it to our design where the \textit{non-transient} stack is only an additional stack to the regular unprotected one.
Based on Intel Pin~\cite{Luk2005pin}, we implemented our own plugin to trace all memory accesses. 
With the plugin, we evaluated how much memory the \textit{non-transient} stack consumes.
For this purpose we ran the GNU Core Utilities, once compiled with the unmodified compiler, and once compiled with our extended LLVM compiler. 
Even for these lightweight applications, we measured a reduction of average \textit{non-transient} stack memory by \SI{42.74}{\percent}. 
The modified LLVM compiler sustained an average \textit{non-transient} stack usage of \SI{4.7}{\kilo\byte}, whereas the applications compiled with a vanilla compiler consumed on average \SI{8.2}{\kilo\byte} on the single \textit{non-transient} stack. 
Moreover, for 64 out of the 91 tested applications (\ie \SI{70.3}{\percent}), the compiler extension reduced the \textit{non-transient} stack usage to only \SI{3528}{\byte}, which is below the smallest memory region that can be set \textit{non-transient}, \ie the size of one virtual page (\SI{4}{\kilo\byte}). 
The reason for these reductions is that the stack is not used anymore for storing user-defined variables.
Hence, the compiler extension makes it practical to deploy \DefenseName with the additional \textit{non-transient} stack.

\paragraph{\DefenseNameLight}
We evaluated the performance impact of \DefenseNameLight, both for unmodified applications as well as applications where we annotate secret values as such. 
For unmodified applications, we do not expect any runtime overhead, except for a constant initialization overhead.

We confirmed this assumption experimentally. 
The average initialization overhead when starting an application with our current non-optimized implementation is \SI{0.15}{\milli\second}. 

For applications with annotated secret values, there is a performance overhead for architectural accesses to the secret. 
Without \DefenseNameLight, the secret could be stored in L1, L2, or L3 cache, or in the main memory.
Hence, the maximum overhead for a memory access is the difference between an L1 cache hit and a cache miss. 
The minimum overhead for a memory access is zero (\ie cache miss in both cases). 
In practice we often see a cache miss instead of an L3 cache hit, which makes an average overhead of \SIx{100} cycles on our test system. 
We evaluated the performance by encrypting a message using OpenSSL's RSA. 
We verified that indeed all memory allocations in OpenSSL use the secure functions using \texttt{ltrace} and single-stepping.
The performance overhead we measured when annotating all buffers that may (temporarily) contain secrets in an RSA encryption is \SI{71.14}{\percent} ($\pm$ \SI{4.66}{\percent}, $n=10\,000$).
This is not surprising as RSA performs many in-place operations in one secure buffer, and hence, higher overheads are expected.
However, this overhead is in the same range as the overhead of the recommended mitigation strategy for Spectre-PHT attacks alone, \ie \SIx{62}--\SI{74.8}{\percent} for serialization barriers~\cite{Carruth2018Hardening}. Aditional overheads are caused by Spectre-BTB mitigations, \eg STIBP (\SIx{30}--\SI{50}{\percent})~\cite{Larabel2018stibp} and IBRS (\SIx{20}--\SI{30}{\percent})~\cite{Tkachenko2018ibrs_performance}, as well as mitigations for other Spectre and Meltdown variants.
Hence, \DefenseName is a viable alternative as its overhead are inherently lower than the ones we observe with \DefenseNameLight, and \DefenseNameLight already is in the range of state-of-the-art mitigation approaches.
\DefenseName improves the performance of \DefenseNameLight by regular caching and by hiding the latency of register loads, hence, the performance will be significantly higher.

\section{Discussion}\label{sec:discussion} 
\DefenseName is not a defense for commodity systems.
\DefenseName requires changes across all layers.
Yet, compared to all other defenses, it is the first proposal to achieve complete protection~\cite{Miller2018taxonomy,Canella2018}.
Concurrent to our work NVIDIA patented a similar idea~\cite{Boggs2019memory}
However, they focus solely on the protection of memory locations, \ie not speculating on memory that might contain secrets.
In contrast to their work, we do provide protection on a register-level, allowing speculatively cache and register fills.
This clearly has a lower performance impact.
However, the various patents in this area~\cite{Intel2014TaintTracking,VMWare2013TaintTracking,Boggs2019memory} give us additional confidence of the practicality of our approach.

Naturally, \DefenseName is particularly interesting in cases where isolation is not clear, \eg to protect a sandbox environment from the sandboxed code.
There are different ways to select what are secrets to protect.
One extreme would be to generally mark all data secret.
As this is not practical related works either restrict it to an architecturally already defined group, or let the user annotate secrets.
Taram~\etal\cite{Taram2019} defined all userspace memory and user input as secret.
However, this can be very expensive, and consequently, Yu~\etal\cite{Yu2019data} proposed a less expensive annotation-based protection mechanism.
While this is an important discussion it is orthogonal to this work.
Our work shows that if we can mark secrets, we can provide complete protection.
From a problem which is, according to Mcilroy~\etal\cite{Mcilroy2019}, currently not solvable in software, \DefenseName shifts the landscape such that the problem is not \emph{easy} to solve, but \emph{solvable} in software.

\paragraph{Dealing with Edge Cases}
As we described in \cref{sec:limitations}, \DefenseName does not support tainting registers which are used to steer the control flow, \eg the instruction pointer or the flags register. 
If the control flow depends on the secret, the code is inherently not side-channel resilient, \ie other side channels such as cache attacks can already be used to extract the secret. 
Hence, we consider only cases where the secret is not leaked via the control flow.
In this case, the commonality of all remaining transient-execution attacks is that the secret moves through a register.
\DefenseName does not prevent any operations from loading data into registers, but it prevents values from being passed on from tainted registers.

There are many elements in a processor that generally could leak data such that a register contains a secret.
No matter whether the data was leaked from---the memory, the cache, the line fill buffer, the load buffer, the store buffer, or just another register---if the register is tainted, \DefenseName does not execute any operation that depends the value from that register.
Hence, under the assumption that the secret has to move through a register (or already be in a register), the protection \DefenseName provides is complete.
Only violating this assumption would allow bypassing \DefenseName.
To the best of our knowledge there is no mechanism on x86-64 that would allow performing \eg an indexed array access without loading the index into a register.
This supports our assumption.

As \DefenseName prevents the value from being passed on from the tainted register, we do not have any edge cases around the various microarchitectural elements.

\paragraph{Microcode}
\DefenseName likely cannot be implemented (efficiently) in microcode or microcode updates.
The reason is that the behavior in the critical path when forwarding a value from a register to a dependent instruction has to be modified.
To the best of our knowledge, there is no microcode involved in this part for performance reasons.

\paragraph{Virtualization}
Our approach is oblivious to virtualization.
EPTs equally contain \textit{non-transient} bits.
Identical to the way several other page table bits are combined (\eg the \textit{non-executable} bit), if any bit in the hierarchy is set to \textit{non-transient}, the page is \textit{non-transient}.
Naturally, the extensions we implemented on the operating system level would have to be identically implemented on the hypervisor level.
We leave this implementation effort for future work.

\paragraph{Implementation of the Microarchitectural Changes}
While a microarchitectural implementation would be interesting, this is not necessary to see the practicality of our work.
We already have the uncacheable memory mapping which are marked in the page table.
Uncacheable memory is not used during speculative execution, although if it is already in a cache, line fill buffer, load buffer, or store buffer, it might be leaked.
Hence, there is already a mechanism in current processors which is very similar to the one we propose.
While uncacheable memory is much slower than what we propose, it clearly shows that an implementation is possible.

\section{Conclusion}\label{sec:conclusion} 
In this paper, we presented \DefenseName, a technique to effectively and efficiently prevent the leakage of secrets during transient execution.
The basic idea of \DefenseName is to transform Spectre from a problem that cannot be solved purely in software~\cite{Mcilroy2019}, to a problem that is not easy to solve, but solvable in software.
For this, \DefenseName requires minimal modifications of applications, compilers, operating systems, and the hardware.
We implemented these in applications, compilers, and operating systems, as well as in a processor simulator.

Mitigating all transient execution attacks with a principled approach of course costs performance.
We provide an approximate proof-of-concept for \DefenseName which we use on commodity systems to obtain a loose upper bound for the performance overhead.
As seen in our security evaluation, \DefenseName is a first proposal for a principled defense tackling the root cause of transient execution attacks.
\DefenseName has no performance overhead for regular applications and even with the over-approximation of \DefenseNameLight, \SI{71.14}{\percent} for security-critical applications, which below the combined overhead of recommended state-of-the-art mitigation strategies.
The overhead with \DefenseName is below \SI{1}{\percent} for most real-world workloads.
Our work shows that transient execution can be made secure while maintaining a high system performance.

\section*{Acknowledgments}
We want to thank Jon Masters for discussions on Spectre mitigations. Concurrent to our research, Jon Masters independently published a brief sketch of a similar idea.
This project has received funding from the European Research Council (ERC) under the European Union's Horizon 2020 research and innovation programme (grant agreement No 681402).
It was also supported by the Austrian Research Promotion Agency (FFG) via the project ESPRESSO, which is funded by the province of Styria and the Business Promotion Agencies of Styria and Carinthia.
It was also supported by the Austrian Research Promotion Agency (FFG) via the K-project DeSSnet, which is funded in the context of COMET – Competence Centers for Excellent Technologies by BMVIT, BMWFW, Styria and Carinthia.
It was also supported by the Austrian Research Promotion Agency (FFG) via the competence center Know-Center (grant number 844595), which is funded in the context of COMET – Competence Centers for Excellent Technologies by BMVIT, BMWFW, and Styria.
Additional funding was provided by a generous gift from Intel and a generous gift from ARM.
Any opinions, findings, and conclusions or recommendations expressed in this paper are those of the authors and do not necessarily reflect the views of the funding parties.

{\footnotesize \bibliographystyle{acm-url}
\bibliography{main}}

\end{document}